\documentclass[nofootinbib,aps,pra,showpacs,superscriptaddress,preprint]{revtex4}

\usepackage{amsmath}

\usepackage{subfig}

\usepackage{graphicx}

\catcode`ð=\active
\defð{\u{g}}
\catcode`Ð=\active
\defÐ{\u{G}}
\catcode`Ý=\active
\defÝ{\. I}
\catcode`ö=\active
\defö{\"{o}}
\catcode`Ö=\active
\defÖ{\"O}
\catcode`ü=\active
\defü{\"{u}}
\catcode`Ü=\active
\defÜ{\"{U}}
\catcode`Þ=\active
\defÞ{\c{S}}
\catcode`þ=\active
\defþ{\c{s}}
\catcode`ý=\active
\defý{{\i}}
\catcode`ç=\active
\defç{\d{c}}
\catcode`Ç=\active
\defÇ{\d{C}}

\begin{document}

\title{Scattering and Bound State Solutions of Asymmetric Hulth\'{e}n Potential}

\author{\small Altuð Arda}
\email[E-mail: ]{arda@hacettepe.edu.tr}\affiliation{Department of
Physics Education, Hacettepe University, 06800, Ankara,Turkey}
\author{\small Oktay Aydoðdu}
\email[E-mail: ]{oktaydogdu@gmail.com}\affiliation{Department of
Physics, Mersin University, 33343, Mersin,Turkey}
\author{\small Ramazan Sever}
\email[E-mail: ]{sever@metu.edu.tr}\affiliation{Department of
Physics, Middle East Technical  University, 06531, Ankara,Turkey}

\begin{abstract}
One-dimensional time-independent Schrödinger equation is solved
for the asymmetric Hulth\'{e}n potential. Reflection and
transmission coefficients and bound state solutions are obtained
in terms of the hypergeometric functions. It is observed that the
unitary condition is satisfied in non-relativistic region.
\end{abstract}

\maketitle

\newpage

\section{Introduction}
The solutions including scattering and/or bound states of the wave
equations have been great interest in quantum mechanical systems
[1-22]. To achieve full information about the system under
consideration one has to investigate the bound as well as
scattering state problem. In Ref. [3], the authors have obtained
the analytical scattering state solutions of the $\ell$-wave
Schrödinger equation for the Eckart potential. $\ell$-wave
continuum states of the Schrödinger equation for the modified
Morse potential have been studied by Wei \textit{et.al.} [7] where
they have obtained the normalized analytical radial wave functions
and derived a corresponding calculation of phase shifts. Chen
\textit{et.al.} have found the exact solutions of scattering
states for the $s$-wave Schrödinger equation with the
Manning-Rosen potential by using standard method [4]. In view of a
spatially one-dimensional Woods-Saxon potential, the scattering
solutions of the Klein-Gordon equation have been obtained in terms
of hypergeometric functions by Rojas \textit{et.al.} and they have
derived the condition for the existence of transmission resonances
[10]. In an arbitrary dimension, Chen \textit{et.al.} have
presented the properties of scattering state solutions of the
Klein-Gordon equation for a Coulomb-like scalar plus vector
potentials [16]. In Ref. [19], low-momentum scattering in the
Dirac equation have been studied. Villalba and Greiner [17] have
investigated the transmission resonances and supercritical states
by solving the two-component Dirac equation for the cusp
potential. In this manner, we intend to search the transmission
and reflection coefficients and eigenvalues of the one-dimensional
Schrödinger equation for the asymmetric Hulth\'{e}n potential
(ASHp).

The "usual" Hulth\'{e}n potential [23] is one of the significant
exponential potential which behaves like a Coulomb potential for
small values of spatially coordinate. The Hulth\'{e}n potential
has many application areas in physics such as atomic physics [24,
25], nuclear and high energy physics [26], solid state physics
[27] and chemical physics [28]. In addition, the Hulth\'{e}n
potential and its various forms are used in relativistic and
non-relativistic regions [8, 9, 12, 13, 14]. In Ref. [8], the
approximate analytical scattering state solutions of the
Schrödinger equation with the generalized Hulth\'{e}n potential
for  any $\ell$-state have been obtained. Saad [9] has studied the
bound states of a spin-$0$ particle in $D$-dimensions and found
the normalization constant in terms of incomplete Beta function.
The scattering solutions of the Klein-Gordon equation for the
general Hulth\'{e}n potential have been obtained and transmission
resonances investigated in Ref. [13]. Guo \textit{et.al.} [21]
have found the transmission resonances for a Dirac particle in the
presence of the Hulth\'{e}n potential in one-dimension. On the
other hand, solutions of the bound and scattering states of the
wave equations for the asymmetric potentials have been recently
examined [18, 22, 29]. In Ref. [18], the authors have investigated
the low-momentum scattering of a Dirac particle in the presence of
cusp potential. In $(1+1)$-dimensions, transmission resonances in
the Duffin-Kemmer-Petiau (DKP) equation for an asymmetric cusp
potential have also been obtained [29]. Recently, Sogut
\textit{et.al.} have examined the scattering and bound state
solutions of the DKP equation in the presence of the ASHp [22]. In
the present work, we study the scattering and bound state
solutions of the one-dimensional Schrödinger equation for the ASHp
[22]
\begin{eqnarray}
V(x)=V_{0}\left[\theta(-x)\frac{e^{ax}}{1-qe^{ax}}+\theta(x)\frac{e^{bx}}{1-\tilde{q}e^{bx}}\right]\,,
\end{eqnarray}
where $V_{0}$ is the strength of the potential and $a, b, q$ and
$\tilde{q}$ are positive parameters. $\theta(x)$ is the Heaviside
step function and for the parameters $q$ and $\tilde{q}$ hold
$q<1$ and $\tilde{q}<1$. Fig. (1) shows dependence of the ASHp
barrier on these parameters.

The organization of the present work is as follows. In Section 2,
we search the reflection and transmission coefficients in terms of
hypergeometric functions for the ASHp barrier by using the form of
the wave functions for $x \rightarrow \pm\infty$. In Section 3, we
obtain a condition for extracting energy eigenvalue for the ASHp
well. This condition is a transcendental equation which can be
solved numerically. We give some numerical values of the energy
eigenvalues for the bound states for chosen values of the
potential parameters. We summarize our results in Section 4.
\section{Reflection and Transmission Coefficients} The
one-dimensional time-independent Schrödinger equation for a
particle with mass $m$ moving in a potential $V(x)$ reads
\begin{eqnarray}
\left\{\frac{d^2}{dx^2}+2m\left(E-V(x)\right)\right\}\psi(x)=0\,.
\end{eqnarray}
Now we look for the solution of the ASHp barrier for the region
$x<0$. Inserting Eq. (1) into Eq. (2) gives
\begin{eqnarray}
\left\{\frac{d^2}{dx^2}+2m\left[E-\frac{V_{0}}{e^{-\,ax}-q}\right]\right\}\psi_{L}(x)=0\,,
\end{eqnarray}
Using a new variable $y=qe^{ax}$ in Eq. (3) one obtains the
following equation
\begin{eqnarray}
y(1-y)\psi''_{L}(y)+(1-y)\psi'_{L}(y)+\frac{1}{y(1-y)}\left\{\beta_{1}-\beta_{2}y+\beta_{3}y^2\right\}\psi_{L}(y)=0\,,
\end{eqnarray}
where
\begin{eqnarray}
\beta_{1}=\frac{2mE}{a^2}\,;\,\,\,\beta_{2}=\frac{4mE}{a^2}+\frac{2mV_{0}}{qa^2}\,;\,\,\,
\beta_{3}=\frac{2mE}{a^2}+\frac{2mV_{0}}{qa^2}\,.
\end{eqnarray}
Taking the trial wave function
\begin{eqnarray}
\psi_{L}(y)=y^{\mu}(1-y)^{\nu}f(y)\,,
\end{eqnarray}
and inserting it into Eq. (4) we have
\begin{eqnarray}
y(1-y)f''(y)+\left[1+2\mu-(2\mu+2\nu+1)y\right]f'(y)-(\mu+\nu+\gamma)(\mu+\nu-\gamma)f(y)=0\,,
\end{eqnarray}
which has the form of the hypergeometric-type equation [30]
\begin{eqnarray}
s(1-s)\chi''+[\zeta_3-(\zeta_1+\zeta_2+1)s]\chi'-\zeta_1\zeta_2\chi=0\,.
\end{eqnarray}
whose solution is given as $\,_2F_1=(\zeta_1,\zeta_2;\zeta_3;s)$.
So, comparing Eq. (7) with Eq. (8) gives us the solution
\begin{eqnarray}
f(y)=A_{1}\,_{2}F_{1}(\mu+\nu-\gamma,\mu+\nu+\gamma;1+2\mu;y)\nonumber\\+
A_{2}y^{-2\mu}\,_{2}F_{1}(-\mu+\nu-\gamma,-\mu+\nu+\gamma;1-2\mu;y)\,,
\end{eqnarray}
and the whole solution for the region $x<0$
\begin{eqnarray}
\psi_{L}(y)=A_{1}y^{\mu}(1-y)^{\nu}\,_{2}F_{1}(\mu+\nu-\gamma,\mu+\nu+\gamma;1+2\mu;y)\nonumber\\+
A_{2}y^{-\mu}(1-y)^{\nu}\,_{2}F_{1}(-\mu+\nu-\gamma,-\mu+\nu+\gamma;1-2\mu;y)\,,
\end{eqnarray}
where
\begin{eqnarray}
\mu=i\frac{k}{a}\,;\,\,\,k=\sqrt{2mE\,}\,;\,\,\,\nu=1\,;\,\,\,\gamma=\frac{i}{a}
\sqrt{2m\left(E+\frac{V_{0}}{q}\right)\,}\,.
\end{eqnarray}
We have to obtain the asymptotic form of the above wave function
since we search the reflection and transmission coefficients. As
$x \rightarrow -\infty$, $y \rightarrow 0$ and $(1-y)^{\nu}
\rightarrow 1$, we obtain from Eq. (10)
\begin{eqnarray}
\psi_{L}(x \rightarrow -\infty)\sim A_{1}q^{\mu}e^{a\mu
x}+A_{2}q^{-\mu}e^{-\,a\mu x}\sim
A_{1}q^{ik/a}e^{ikx}+A_{2}q^{-ik/a}e^{-ikx}\,,
\end{eqnarray}
where we have used $\,_2F_1=(\zeta_1,\zeta_2;\zeta_3;0)=1$.

To obtain the solution of the ASHp barrier for the region $x>0$ we
insert Eq. (1) into Eq. (2) and get
\begin{eqnarray}
\left\{\frac{d^2}{dx^2}+2m\left[E-\frac{V_{0}}{e^{bx}-\tilde{q}}\right]\right\}\psi_{R}(x)=0\,.
\end{eqnarray}
Defining the new variable $z=\tilde{q}e^{-\,bx}$ gives us
\begin{eqnarray}
z(1-z)\psi''_{R}(z)+(1-z)\psi'_{R}(z)+\frac{1}{z(1-z)}\left\{\tilde{\beta_{1}}-
\tilde{\beta_{2}}z+\tilde{\beta_{3}}z^2\right\}\psi_{R}(z)=0\,,
\end{eqnarray}
where
\begin{eqnarray}
\tilde{\beta_{1}}=\frac{2mE}{b^2}\,;\,\,\,\tilde{\beta_{2}}=\frac{4mE}{b^2}+\frac{2mV_{0}}{\tilde{q}b^2}\,;\,\,\,
\tilde{\beta_{3}}=\frac{2mE}{b^2}+\frac{2mV_{0}}{\tilde{q}b^2}\,.
\end{eqnarray}
By using a trial wave function
$\psi_{R}(z)=z^{\mu_{1}}(1-z)^{\nu_{1}}h(z)$ in Eq. (14) we obtain
the whole solution of the ASHp for the region $x>0$
\begin{eqnarray}
\psi_{R}(z)=A_{3}z^{\mu_{1}}(1-z)^{\nu_{1}}\,_{2}F_{1}(\mu_{1}+\nu_{1}-
\gamma_{1},\mu_{1}+\nu_{1}+\gamma_{1};1+2\mu_{1};z)\nonumber\\+
A_{4}z^{-\mu_{1}}(1-z)^{\nu_{1}}\,_{2}F_{1}(-\mu_{1}+\nu_{1}-\gamma_{1},-\mu_{1}
+\nu_{1}+\gamma_{1};1-2\mu_{1};z)\,,
\end{eqnarray}
where
\begin{eqnarray}
\mu_{1}=i\frac{k}{b}\,;\,\,\,k=\sqrt{2mE\,}\,;\,\,\,\nu_{1}=1\,;\,\,\,
\gamma_{1}=\frac{i}{b}\sqrt{2m\left(E+\frac{V_{0}}{\tilde{q}}\right)\,}\,.
\end{eqnarray}
In order to define a plane wave travelling from left to right we
have to set $A_{3}=0$ in Eq. (16), so
\begin{eqnarray}
\psi_{R}(z)=A_{4}z^{-\mu_{1}}(1-z)^{\nu_{1}}\,_{2}F_{1}(-\mu_{1}+\nu_{1}
-\gamma_{1},-\mu_{1}+\nu_{1}+\gamma_{1};1-2\mu_{1};z)\,.
\end{eqnarray}
Now we give the form of the wave function at $x \rightarrow
+\infty$ for region $x>0$. As $x \rightarrow +\infty$, $z
\rightarrow 0$ and $(1-z)^{\nu_{1}} \rightarrow 1$, we have from
Eq. (18)
\begin{eqnarray}
\psi_{R}(x \rightarrow +\infty)\sim
A_{4}(\tilde{q})^{-\mu_{1}}e^{b\mu_{1} x}\sim
A_{4}(\tilde{q})^{-ik/b}e^{ikx}\,.
\end{eqnarray}
As a result we can summarize the wave function for the limit $x
\rightarrow \pm\infty$ from Eq. (12) and Eq. (19) as
\begin{eqnarray}
\psi(x)=\left\{
\begin{array}{ll}
A_{1}q^{ik/a}e^{ikx}+A_{2}q^{-ik/a}e^{-ikx} & x \rightarrow -\infty,\\
A_{4}(\tilde{q})^{-ik/b}e^{ikx} & x\rightarrow +\infty.\\
\end{array}\right.
\end{eqnarray}
The wave function in Eq. (10) can be written as
$\psi_L=\psi_{inc}+\psi_{ref}$ in the limit $x \rightarrow
-\infty$ where $\psi_{inc}$ is the incident and $\psi_{ref}$ is
the reflected wave. Similarly, as $x \rightarrow +\infty$ the wave
function in Eq. (18) is $\psi_R=\psi_{trans}$ where $\psi_{trans}$
is the transmitted wave. These definitions give us the reflection
and transmission coefficients as
\begin{eqnarray}
R&=&\left|\frac{\psi_{ref}}{\psi_{inc}}\right|^2=\frac{\left|A_{2}\right|^2}{\left|A_{1}\right|^2}\,,\nonumber\\
T&=&\left|\frac{\psi_{trans}}{\psi_{inc}}\right|^2=\frac{\left|A_{4}\right|^2}{\left|A_{1}\right|^2}\,.
\end{eqnarray}
In order to give the explicit expressions for the coefficients
used in the above equations we need to use the continuity
conditions on the wave function given as
$\psi_{R}(x=0)=\psi_{L}(x=0)$ and $\psi'_{R}(x=0)=\psi'_{L}(x=0)$
where prime denotes derivative with respect to $x$ . The matching
of the wave functions at $x=0$ gives
\begin{eqnarray}
A_{1}C_{1}F_{1}+A_{2}C_{2}F_{2}=A_{4}C_{3}F_{3}\,,
\end{eqnarray}
and the matching of derivatives of the wave functions reads
\begin{eqnarray}
aqA_{1}C_{1}(D_1F_{1}+D_{4}F_{4})+aqA_{2}C_{2}(D_{2}F_{2}+D_{5}F_{5})\nonumber\\
=b\tilde{q}A_{4}C_{3}(D_{3}F_{3}-D_{6}F_{6})\,.
\end{eqnarray}
where we have used the property of the hypergeometric functions as
$\frac{d}{ds}\,_2F_1(\zeta_1,\zeta_2;\zeta_3;s)=\frac{\zeta_1\zeta_2}{\zeta_3}\,_2F_{1}(\zeta_1+1,\zeta_2+1;\zeta_1+1;s)$.

Combining last two equations we obtain the followings for the
coefficients written in Eq. (21)
\begin{eqnarray}
\frac{A_{2}}{A_{1}}=\frac{C_{1}\left[b\tilde{q}F_{1}(D_{3}F_{3}-D_{6}F_{6})-aqF_{3}(D_1F_{1}+D_{4}F_{4})\right]}
{C_{2}\left[aqF_{3}(D_{2}F_{2}+D_{5}F_{5})-b\tilde{q}F_{2}(D_{3}F_{3}-D_{6}F_{6})\right]}\,,\\
\frac{A_{4}}{A_{1}}=\frac{aqC_{1}\left[F_{1}(D_{2}F_{2}+D_{5}F_{5})-F_{2}(D_{1}F_{1}+D_{4}F_{4})\right]}
{C_{3}\left[aqF_{3}(D_{2}F_{2}+D_{5}F_{5})-b\tilde{q}F_{2}(D_{3}F_{3}-D_{6}F_{6})\right]}\,.
\end{eqnarray}
where the following abbreviations in the above equations have been
used
\begin{eqnarray}
C_{1}&=&q^{\mu}(1-q)^{\nu}\,;\,\,C_{2}=q^{-\mu}(1-q)^{\nu}\,;\,\,
C_{3}=(\tilde{q})^{-\mu_{1}}(1-\tilde{q})^{\nu_{1}}\,,\\
D_{1}&=&\frac{\mu}{q}-\frac{\nu}{1-q}\,;\,\,D_{2}=-\frac{\mu}{q}
-\frac{\nu}{1-q}\,;\,\,D_{3}=\frac{\mu_{1}}{\tilde{q}}+\frac{\nu_{1}}{1-\tilde{q}}\,,\nonumber\\
D_{4}&=&\frac{(\mu+\nu-\gamma)(\mu+\nu+\gamma)}{1+2\mu}\,;
D_{5}=\frac{(-\mu+\nu-\gamma)(-\mu+\nu+\gamma)}{1-2\mu}\,;\nonumber\\
D_{6}&=&\frac{(-\mu_{1}+\nu_{1}-\gamma_{1})(-\mu_{1}+\nu_{1}+\gamma_{1})}{1-2\mu_{1}}\,,\\
F_{1}(\mu,\nu,\gamma,q)&=&\,_2F_{1}(\mu+\nu-\gamma,\mu+\nu+\gamma;1+2\mu;q)\,,\nonumber\\
F_{2}(\mu,\nu,\gamma,q)&=&\,_2F_{1}(-\mu+\nu-\gamma,-\mu+\nu+\gamma;1-2\mu;q)\,,\nonumber\\
F_{3}(\mu_{1},\nu_{1},\gamma_{1},\tilde{q})&=&\,_2F_{1}(-\mu_{1}+\nu_{1}-\gamma_{1},-\mu_{1}
+\nu_{1}+\gamma_{1};1-2\mu_{1};\tilde{q})\,,\nonumber\\
F_{4}(\mu,\nu,\gamma,q)&=&\,_2F_{1}(\mu+\nu-\gamma+1,\mu+\nu+\gamma+1;2+2\mu;q)\,,\nonumber\\
F_{5}(\mu,\nu,\gamma,q)&=&\,_2F_{1}(-\mu+\nu-\gamma+1,-\mu+\nu+\gamma+1;2-2\mu;q)\,,\nonumber\\
F_{6}(\mu_{1},\nu_{1},\gamma_{1},\tilde{q})&=&\,_2F_{1}(-\mu_{1}+\nu_{1}-\gamma_{1}+1
,-\mu_{1}+\nu_{1}+\gamma_{1}+1;2-2\mu_{1};\tilde{q})\,.
\end{eqnarray}
Inserting Eqs. (26), (27) and (28) into Eq. (21) gives the
explicit expressions of the transmission and reflection
coefficients. Figs. (2)-(4) show different variations of these
coefficients according to the energy and also potential strength
for various potential parameter values. It is seen in Fig. (2)
that the unitarity condition, $R+T=1$, is certainly satisfied.
Fig. (3) and left panel of Fig. (4) show that the dependence of
the transmission coefficient on the potential parameters is very
similar which means that it goes to zero for relatively lower
values of energy while goes to unity for higher values of energy.
The right panel of Fig. (4) displays the dependence of the
transmission coefficient on the strength of the potential.
According to this plot, the transmission probability of the
particle from the barrier is exactly one if the height of the
potential is zero as expected. This probability goes to zero with
increasing value of the strength of the potential.
\section{Energy Eigenvalues}
In this section, we deal with the bound state solutions of the
ASHp well which means that $V_0 \rightarrow -V_0$. The equation
(3) for $x<0$ turns into
\begin{eqnarray}
\left\{\frac{d^2}{dx^2}+2m\left[E+\frac{V_{0}}{e^{-\,ax}-q}\right]\right\}\psi(x)=0\,,
\end{eqnarray}
Using the transformation $y=qe^{ax}$ and taking the trial
wavefunction $\psi(y)=y^{\mu_2}(1-y)^{\nu_2}g(y)$, the solution of
Eq. (29) becomes
\begin{eqnarray}
g(y)=A_{5}\,_{2}F_{1}(\mu_{2}+\nu_{2}-\gamma_{2},\mu_{2}+\nu_{2}+\gamma_{2};1+2\mu_{2};y)\nonumber\\+
A_{6}y^{-2\mu_{2}}\,_{2}F_{1}(-\mu_{2}+\nu_{2}-\gamma_{2},-\mu_{2}+\nu_{2}+\gamma_{2};1-2\mu_{2};y)\,,
\end{eqnarray}
with the parameters
\begin{eqnarray}
\mu_{2}=\frac{i}{a}\sqrt{2mE\,}=\mu\,;\,\,\,\nu_{2}=\nu=1\,;\,\,\,\gamma_{2}=\frac{i}{a}
\sqrt{2m\left(E-\frac{V_{0}}{q}\right)\,}\,.
\end{eqnarray}
and the complete solution of Eq. (29) is given
\begin{eqnarray}
\psi_{L}(y)=A_{5}y^{\mu_{2}}(1-y)^{\nu_{2}}\,_{2}F_{1}(\mu_{2}+\nu_{2}-\gamma_{2},
\mu_{2}+\nu_{2}+\gamma_{2};1+2\mu_{2};y)\nonumber\\+
A_{6}y^{-\mu_{2}}(1-y)^{\nu_{2}}\,_{2}F_{1}(-\mu_{2}+\nu_{2}-\gamma_{2},-\mu_{2}+\nu_{2}+\gamma_{2};1-2\mu_{2};y)\,.
\end{eqnarray}

Next, we search the solutions of the following form of Eq. (13)
for $x>0$
\begin{eqnarray}
\left\{\frac{d^2}{dx^2}+2m\left[E+\frac{V_{0}}{e^{bx}-\tilde{q}}\right]\right\}\psi(x)=0\,,
\end{eqnarray}
By using the variable $z=\tilde{q}e^{-\,bx}$ and putting
$\psi(z)=z^{\mu_3}(1-z)^{\nu_3}\omega(z)$ in Eq. (33) we obtain
\begin{eqnarray}
\omega(z)=A_{7}\,_{2}F_{1}(\mu_{3}+\nu_{3}-\gamma_{3},\mu_{3}+\nu_{3}+\gamma_{3};1+2\mu_{3};z)\nonumber\\+
A_{8}z^{-2\mu_{3}}\,_{2}F_{1}(-\mu_{3}+\nu_{3}-\gamma_{3},-\mu_{3}+\nu_{3}+\gamma_{3};1-2\mu_{3};z)\,,
\end{eqnarray}
with the parameters
\begin{eqnarray}
\mu_{3}=\frac{i}{b}\sqrt{2mE\,}=\mu_{1}\,;\,\,\,\nu_{3}=\nu_{1}=1\,;\,\,\,\gamma_{3}=\frac{i}{b}
\sqrt{2m\left(E-\frac{V_{0}}{\tilde{q}}\right)\,}\,.
\end{eqnarray}
Finally, we obtain the complete bound state solution of the
Schrödinger equation for $x>0$
\begin{eqnarray}
\psi_{R}(z)=A_{7}z^{\mu_{3}}(1-z)^{\nu_{3}}\,_{2}F_{1}(\mu_{3}+\nu_{3}-\gamma_{3},
\mu_{3}+\nu_{3}+\gamma_{3};1+2\mu_{3};z)\nonumber\\+
A_{8}z^{-\mu_{3}}(1-z)^{\nu_{3}}\,_{2}F_{1}(-\mu_{3}+\nu_{3}-\gamma_{3},-\mu_{3}+\nu_{3}+\gamma_{3};1-2\mu_{3};z)\,.
\end{eqnarray}
In order to represent the wavefunctions in Eq. (32) and Eq. (36)
of the bound state solutions they satisfy the boundary condition
being zero at infinity which gives $A_6=A_8=0$ and we obtain
\begin{eqnarray}
\psi_{L}(y)\sim
A_{5}y^{\mu_{2}}(1-y)^{\nu_{2}}\,_{2}F_{1}(\mu_{2}+\nu_{2}-\gamma_{2},
\mu_{2}+\nu_{2}+\gamma_{2};1+2\mu_{2};y)\,,\\
\psi_{R}(z)\sim
A_{7}z^{\mu_{3}}(1-z)^{\nu_{3}}\,_{2}F_{1}(\mu_{3}+\nu_{3}-\gamma_{3},
\mu_{3}+\nu_{3}+\gamma_{3};1+2\mu_{3};z)\,.
\end{eqnarray}

Matching last two expressions in $x=0$ requiring continuity of the
wavefunction and of its first derivative gives
\begin{subequations}
\begin{align}\label{eq_basics_a}
A_{5}F_{1}(\mu_{2},\nu_{2},\gamma_{2},q)-A_{7}F_{2}(\mu_{3},\nu_{3},\gamma_{3},\tilde{q})=0\,,\\
\label{eq_basics_b}
A_{5}\left\{\left(\frac{\mu_{2}}{q}-\frac{\nu_{2}}{1-q}\right)\,F_{1}(\mu_{2},\nu_{2},\gamma_{2},q)
+F_{3}(\mu_{2},\nu_{2},\gamma_{2},q)\right\}\nonumber\\-
A_{7}\left\{\left(\frac{\mu_{3}}{\tilde{q}}-\frac{\nu_{3}}{1-\tilde{q}}\right)\,F_{2}(\mu_{3},\nu_{3},\gamma_{3},\tilde{q})
+F_{4}(\mu_{3},\nu_{3},\gamma_{3},\tilde{q})\right\}=0\,.
\end{align}
\end{subequations}
where
\begin{eqnarray}
F_{1}(\mu_{2},\nu_{2},\gamma_{2},q)&=&q^{\,\mu_{2}}(1-q)^{\nu_{2}}\,_2F_{1}(\mu_{2}+\nu_{2}-\gamma_{2},\mu_{2}
+\nu_{2}+\gamma_{2};1+2\mu_{2};q)\,,\nonumber\\
F_{2}(\mu_{3},\nu_{3},\gamma_{3},\tilde{q})&=&(\tilde{q})^{\,\mu_{3}}(1-\tilde{q})^{\nu_{3}}
\,_2F_{1}(\mu_{3}+\nu_{3}-\gamma_{3},\mu_{3}+\nu_{3}+\gamma_{3};1+2\mu_{3};\tilde{q})\,,\nonumber\\
F_{3}(\mu_{2},\nu_{2},\gamma_{2},q)&=&q^{\,\mu_{2}}(1-q)^{\nu_{2}}\,
\frac{(\mu_{2}+\nu_{2}-\gamma_{2})((\mu_{2}+\nu_{2}+\gamma_{2})}{1+2\mu_{2}}\nonumber\\&\times&
\,_2F_{1}(\mu_{2}+\nu_{2}-\gamma_{2}+1,\mu_{2}+\nu_{2}+\gamma_{2}+1;2+2\mu_{2};q)\,,\nonumber\\
F_{4}(\mu_{3},\nu_{3},\gamma_{3},\tilde{q})&=&(\tilde{q})^{\,\mu_{3}}(1-\tilde{q})^{\nu_{3}}\,
\frac{(\mu_{3}+\nu_{3}-\gamma_{3})((\mu_{3}+\nu_{3}+\gamma_{3})}{1+2\mu_{3}}\nonumber\\&\times&
\,_2F_{1}(\mu_{3}+\nu_{3}-\gamma_{3}+1,\mu_{3}+\nu_{3}+\gamma_{3}+1;2+2\mu_{3};\tilde{q})\,.
\end{eqnarray}

Equation (39) has a nontrivial solution only if its determinant is
zero. Using this equation, one can determine the energy
eigenvalues of the ASHp well numerically. Here, we give our
numerical results for the energy eigenvalues as a list for some
values of the parameters, for example, $m=1, a=0.5, b=0.75,
V_{0}=5, q=0.1$ and $\tilde{q}=0.5$ taking into account that
$-\left|V_{0}\right|<E<0$: $E_1=-2.453010, E_2=-2.251290,
E_3=-0.924802, E_4=-0.491271, E_5=-0.001356$ (in atomic unit).

\section{Results and Conclusions}
We solve the one-dimensional Schrödinger equation for the
asymmetric Hulth\'{e}n potential. We find the transmission and
reflection coefficients for the ASHp barrier in terms of
hypergeometric functions and give some plots showing the
dependence of these coefficients on the potential parameters $a,
b, q, \tilde{q}, V_{0}$ and on the energy $E$. We observe that the
unitarity condition is exactly satisfied in all cases. We also
compute the energy eigenvalues for the bound states extracting an
eigenvalue equation which can be solved numerically. We calculate
five different energy eigenvalues by taking into account that
$-\left|V_{0}\right|<E<0$.

\section{Acknowledgments}
This research was partially supported by the Scientific and
Technical Research Council of Turkey.

\newpage

\begin{figure}
\centering \subfloat{
\includegraphics[height=2.2in,width=2.8in]{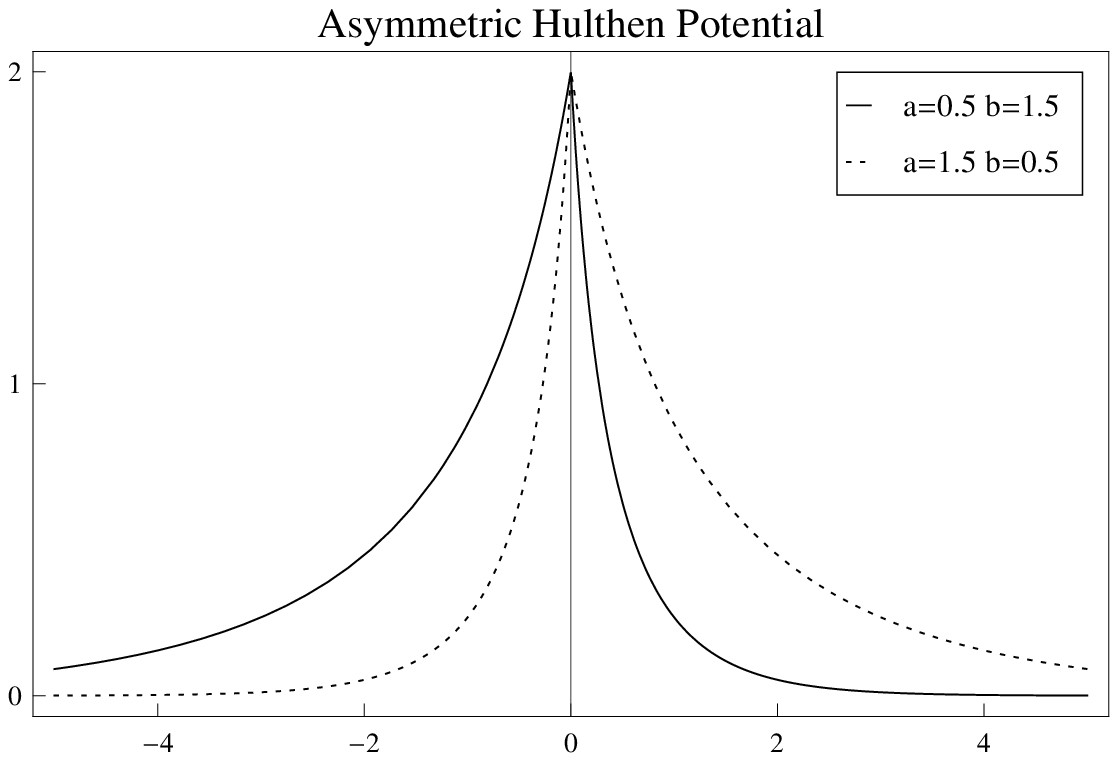}}
\hspace{0.1\linewidth} \subfloat{
\includegraphics[height=2.2in,width=2.8in]{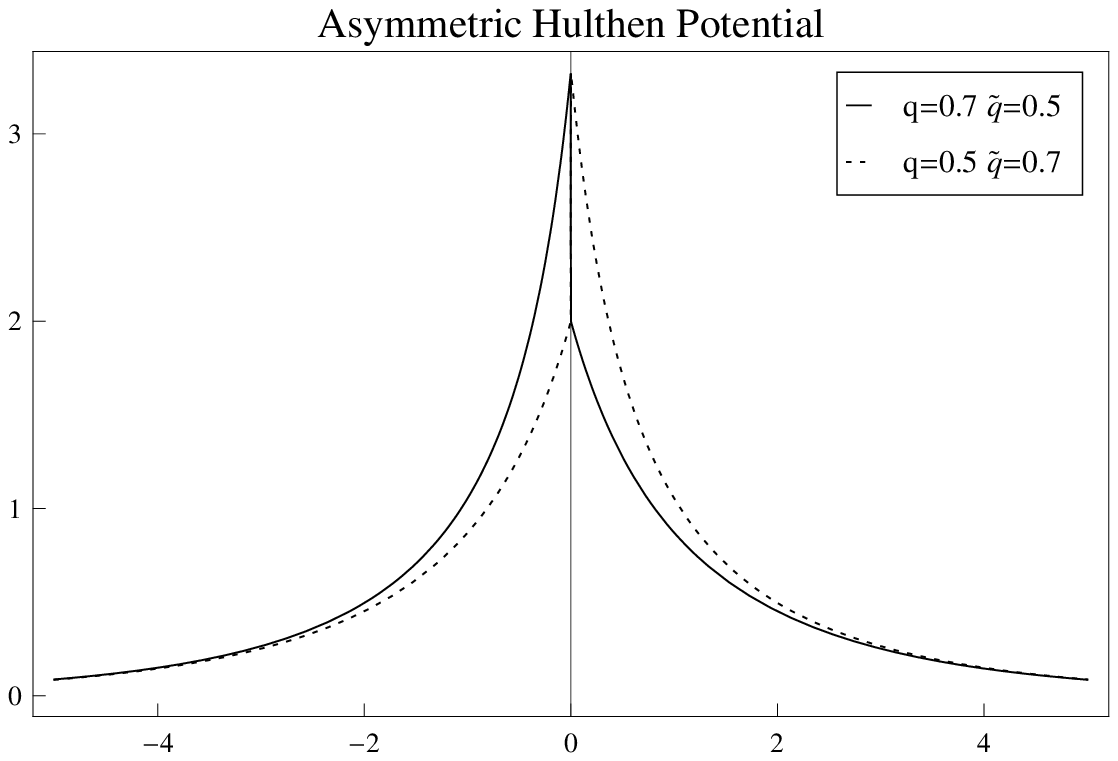}}
\caption{plots of ASHp for different values of the potential
parameters for $V_0=1, q=\tilde{q}=0.5$ (left panel) and $V_0=1,
a=b=0.5$ (right panel).}
\end{figure}

\begin{figure}
\centering
\includegraphics[height=2.5in, width=4in, angle=0]{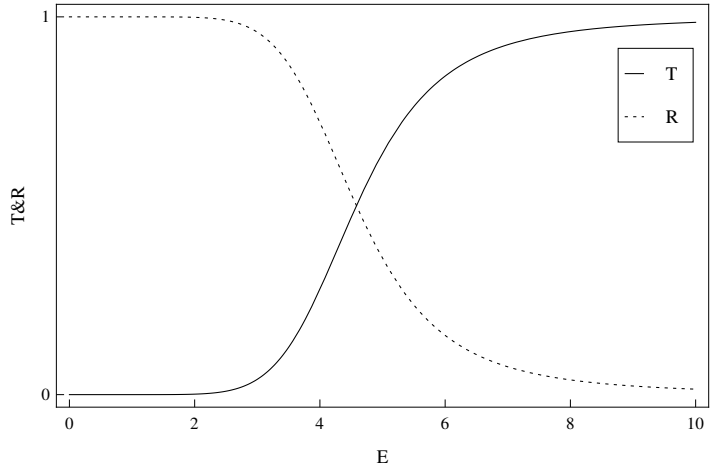}
\caption{transmission ($T$) and reflection ($R$) coefficients
varying with $E$  for $a=0.4, b=0.5, q=0.6, \tilde{q}=0.7, m=1,
V_0=2$.}
\end{figure}

\begin{figure}
\centering \subfloat{
\includegraphics[height=2.2in,width=2.8in]{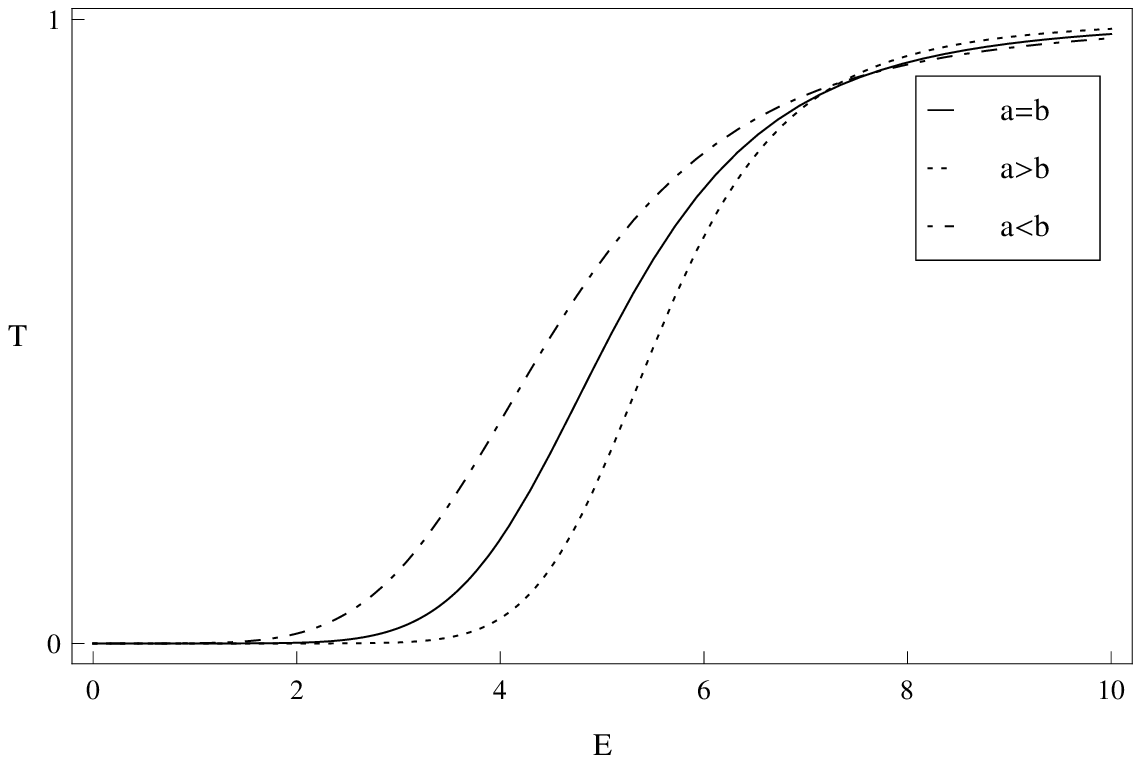}}
\hspace{0.1\linewidth} \subfloat{
\includegraphics[height=2.2in,width=2.8in]{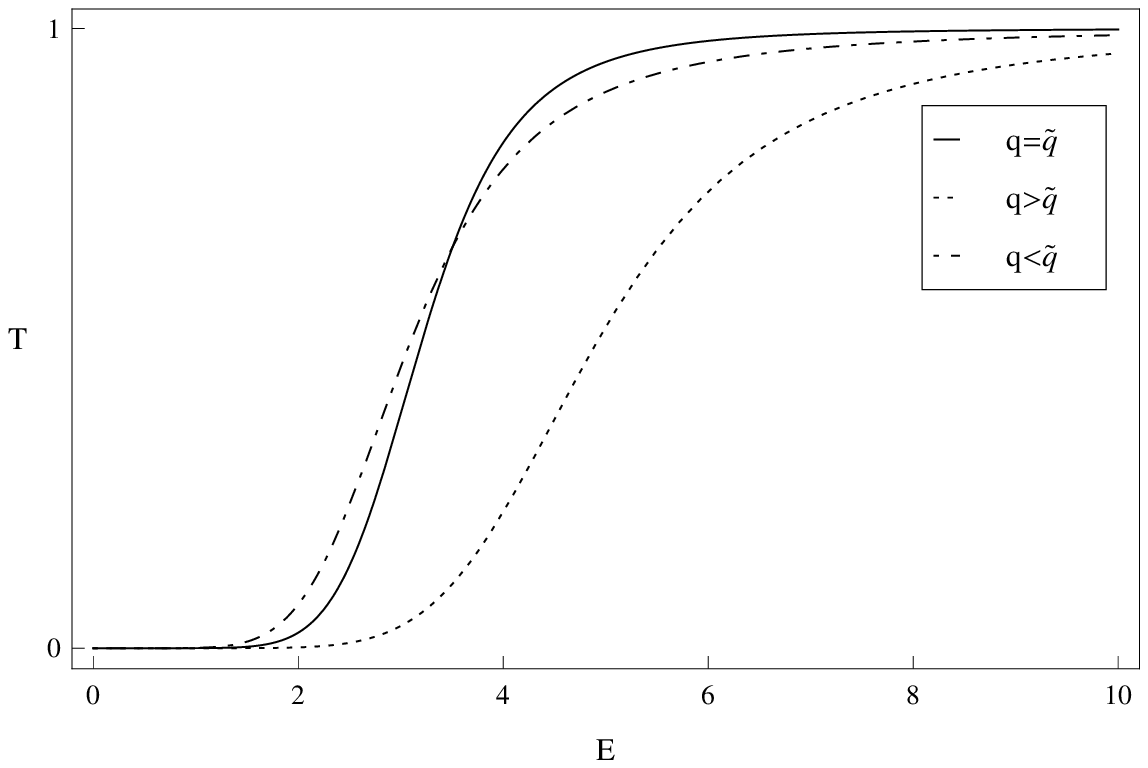}}
\caption{variation of the transmission coefficient with different
potential parameters for $m=1, V_0=2$ (left panel:$a=b=0.5; a=0.8,
b=0.3; a=0.3, b=0.8; q=\tilde{q}=0.7$ right
panel:$q=\tilde{q}=0.5; q=0.6, \tilde{q}=0.4; q=0.4,
\tilde{q}=0.6; a=b=0.5$).}
\end{figure}

\begin{figure}
\centering \subfloat{
\includegraphics[height=2.2in,width=2.8in]{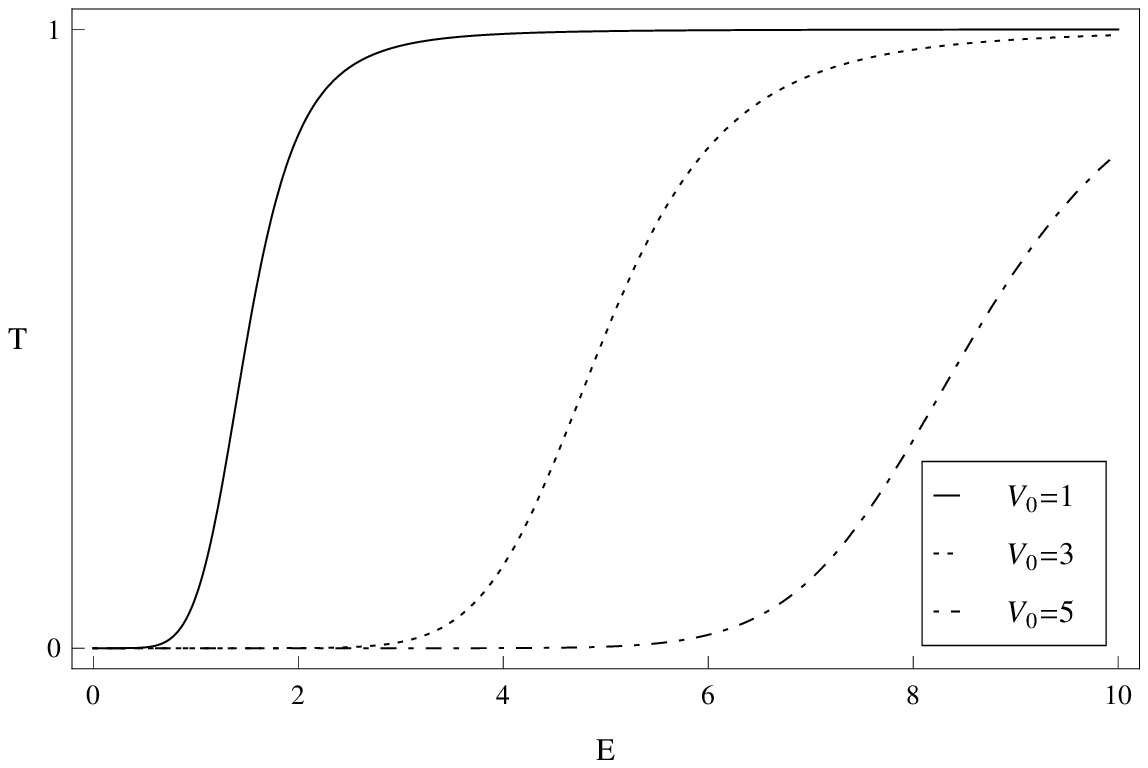}}
\hspace{0.1\linewidth} \subfloat{
\includegraphics[height=2.2in,width=2.8in]{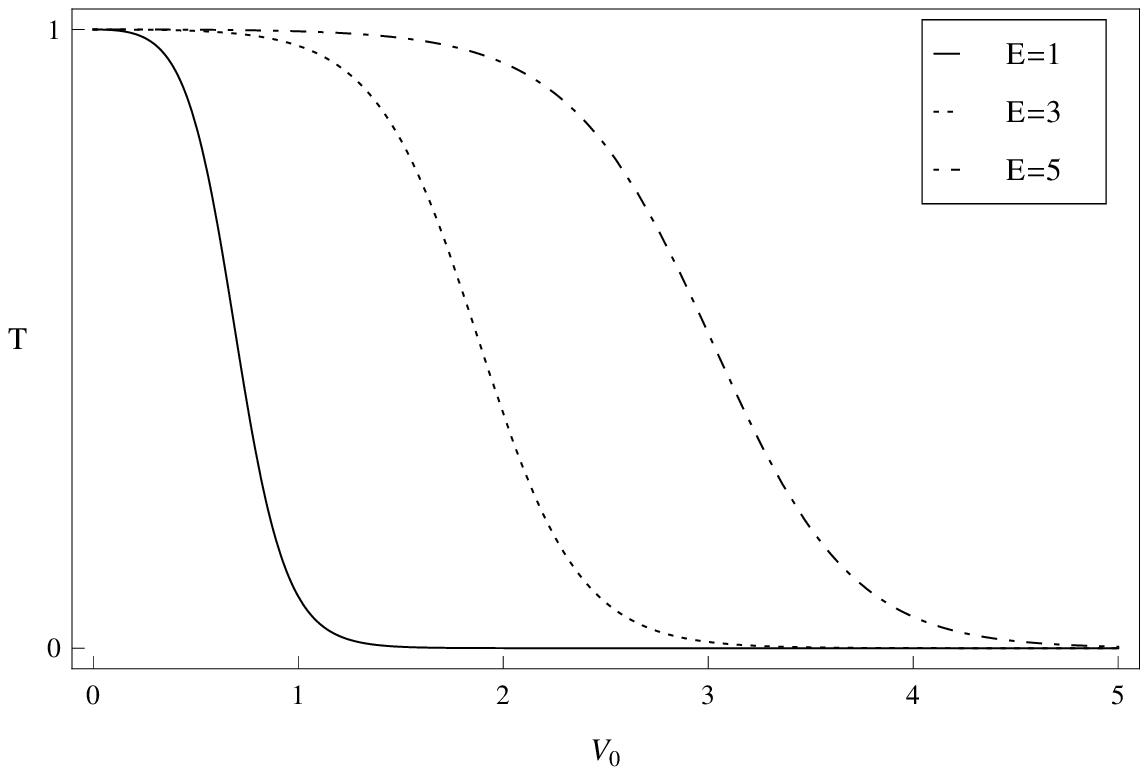}}
\caption{variation of the transmission coefficient with energy $E$
and potential parameter $V_0$ for $a=b=q=\tilde{q}=0.5, m=1$.}
\end{figure}

\end{document}